\documentclass[12pt]{article}

\usepackage{amsmath,amssymb}
\usepackage{pifont,fancybox,shadow}
\usepackage{graphicx}
\usepackage{psfrag}
\usepackage{epsf,rotate}
\usepackage[english]{babel}

\usepackage{epsfig}
\usepackage{rotating}

\begin{document}
\baselineskip = 20pt

\def\gvect#1{\mbox{\boldmath $#1$}}

\newcommand{\TIT}[1]{\begin{center}\shadowbox{\Huge\sc \vbox{#1}}\end{center}}
\newcommand{\heading}[1]{\begin{center}\shadowbox{\Large\bf \vbox{#1}}
\end{center}}


\title{Long term persistence in the sea surface temperature
fluctuations}

\author{Roberto A. Monetti$^{(1)}$ \footnote{Present address: Center for Interdisciplinary Plasma Science (CIPS),
Max-Planck-Institut f\"{u}r extraterrestrische Physik,
Giessenbachstr. 1, 85749 Garching, Germany.}$\,$, Shlomo Havlin$^{(1)}$, and Armin Bunde$^{(2)}$ \\
{\small (1)} Minerva Center and Department of Physics \\ Bar-Ilan University, Ramat-Gan 52900, Israel \\
{\small (2)} Institut f\"{u}r Theoretische Physik III \\ Justus-Liebig-Universit\"{a}t
Giessen \\ Heinrich-Buff-Ring 16, 35392 Giessen, Germany}

\maketitle

\begin{abstract}
We study the temporal correlations in the sea surface
temperature (SST) fluctuations around the seasonal mean values in the
Atlantic and Pacific oceans. We apply a method that
systematically overcome possible trends in the data. We
find that the SST persistence, characterized by
the correlation $C(s)$ of temperature fluctuations separated by a time
period $s$, displays two different regimes. In the short-time regime
which extends up to roughly $10$ months, the temperature fluctuations
display a nonstationary behavior for both oceans, while
in the asymptotic regime it becomes stationary. The long term
correlations decay as $C(s) \sim s^{-\gamma}$ with
$\gamma \sim 0.4$ for both oceans which
is different from $\gamma \sim 0.7$ found
for atmospheric land temperature.
\end{abstract}
\newpage
\section{Introduction}
The oceans cover almost three quarters of the Earth's surface and have the
greatest capacity to store heat. Thus, they are able to
regulate the temperature on land even in sites located far away from
the coastline. This property of the oceans suggests that they may
posses a strong temperature persistence, i.e. a strong tendency that
the water temperature of a particular day will remain the same the
next day. Persistence can be characterized by the auto-correlation
function $C(s)$ of temperature  variations separated by a time period
$s$. Recently, quantitative studies of the persistence in atmospheric
land temperatures revealed that atmospheric land temperature
fluctuations exhibit long-range power law correlations $C(s) \sim
s^{-\gamma}$ with $\gamma \approx 0.70$ [Koscielny-Bunde et al., 1996;
Koscielny-Bunde et al., 1998; Pelletier, 1997; Pelletier et al., 1997;
Talkner et al., 2000]. In this Letter, we study the persistence in sea
surface temperature (SST) records at many sites in the Atlantic and
Pacific oceans using the detrended fluctuation analysis (DFA) method
[Peng et al., 1994; Kantelhardt et al., 2001].
We find that for all time scales the SST fluctuations exhibit stronger
correlations than atmospheric land temperature
fluctuations. The long term persistence of the SST is characterized by
a correlation exponent $\gamma \sim 0.4$ for both oceans.

We have analyzed both types of SST data sets that are available, the
monthly SST
for the period $1856-2001$ and the weekly SST for the period
$1981-2001$.
For the period 1856-1981, the monthly SST data sets were
obtained by
Kaplan et al. [Kaplan et al., 1998] who used optimal
estimation in space applying 80 empirical orthogonal functions
to interpolate ship observations of the United Kingdom Meteorological Office
database [Parker et al., 1994].
After 1981, the monthly data correspond to the
projection of the National
Center for Environmental Prediction (NCEP) optimal interpolation (OI)
analysis [Reynolds et al., 1993;  Reynolds et al., 1994]. The weekly SST's also
correspond to the
NCEP OI analysis. The data are freely available from
http://ingrid.ldeo.columbia.edu/SOURCES/.

\section{The Method}
We focus on the temperature fluctuations around the periodical
seasonal trend. In order to remove the periodical trend, we first
determine the mean temperature $\langle T_a\rangle$ for each month/week by
averaging over all years in the time series. Then, we analyze the
temperature deviations $\Delta T_i = T_i - \langle T_a\rangle $ from
these mean values. The persistence in the temperature fluctuations can
be characterized by the auto-correlation function,
\begin{equation}
C(s) \equiv \langle\Delta T_i \Delta T_{i+s}\rangle  = \frac{1}{N-s}
\sum_{i=1}^{N-s} \Delta T_i \Delta T_{i+s},
\end{equation}
where $N$ is the length of the record and $s$ is the time lag. A
direct calculation of $C(s)$ is hindered by the level of noise present
in the finite temperature series and by possible nonstationarities in
the data. To reduce the noise, we study the temperature profile
function,
\begin{equation}
Y_k = \sum_{i=1}^{k} \Delta T_i.
\end{equation}
We can consider the profile $Y_k$ as the position of a random walker
on a linear chain after $k$ steps. According to the random walk
theory, the fluctuations $F(s)$ of the profile in a given time window
of length $s$ are related to the correlation function $C(s)$.
For the relevant case of long-range power law correlations,
\begin{equation}
C(s) \sim s^{-\gamma}, \hskip 1 true cm 0 < \gamma < 1,
\end{equation}
the fluctuations increase as a power law [Barabasi et al., 1995; Shlesinger et al.,
1987],
\begin{equation}
F(s) \sim s^{\alpha}, \hskip 1 true cm \alpha = 1 - \gamma/2.
\end{equation}
For uncorrelated data ($\gamma \ge 1$), we have $\alpha = 1/2$.
To find how the fluctuations scale with $s$, we divide the profile
into non-overlapping intervals of length $s$. We calculate the square
fluctuations $F^{2}_{\nu}(s)$ in each interval $\nu$ and obtain $F(s)$
by averaging over all intervals, $F(s) \equiv
\langle F^{2}_{\nu}(s)\rangle ^{1/2}$. Here, we use two methods that
differ in the
way fluctuations are measured. In the fluctuation analysis (FA), the
square of the fluctuations is defined as $ F^{2}_{\nu}(s) =
(Y_{(\nu +1)s} - Y_{\nu s})^2$ where $Y_{\nu s}$ and $Y_{(\nu +1)s}$
are the values of the profile at the beginning and the end of each
segment $\nu$, respectively. In
the detrended fluctuation analysis, we determine in each
interval the best polynomial fit of the profile and define
$F_{\nu}(s)$ as the variance between the
profile and the best fit in the intervals. Different orders $n$ of DFA
(DFA1, DFA2, etc)
differ in the order of the polynomial used in the fitting procedure.
By construction, FA is sensitive to any kind of trend and
thus equivalent to the Hurst and the power spectrum analyses. In
contrast, DFA$n$ removes a polynomial trend of order
$n-1$ in the temperature record and thus, it is superior to the conventional
methods.

To characterize the persistence, we have
applied the FA and DFA methods to $36$ ($46$) monthly SST records and $64$
($35$) weekly SST records in the Atlantic (Pacific) ocean.

\section{Results and Discussion}
Figure
1(a-c) show three typical
plots of the monthly
temperature profile function $Y_t$
for a land station (Prague), a site in the Atlantic ocean, and a
site in the Pacific ocean, respectively.
Parabolic-like profile functions which are
concave (convex) may indicate the
presence of a positive (negative) linear trend (see Eq. 2). However,
Fig. 1(d) illustrates that pure correlated data may also lead to
parabolic-like profile functions. Trends and
correlations can be distinguished and characterized by comparing
the FA and DFA
results [Kantelhardt et al., 2001; Govindan et al., 2001]. Figure 2 shows log-log
plots of the FA and
DFA curves for the profiles shown in Fig. 1.
Figure 2(a)
shows that at large times Prague
temperature fluctuations display a power law behavior. The fluctuation
exponent obtained from the FA (0.81) is greater than the values given
by the DFA1-5 (0.65). This difference
is probably due to the effect of the well known urban warming of Prague. The
fluctuation exponent $\alpha \approx 0.65$ is
consistent with the earlier finding, where
the whole Prague record (218 years) has been analyzed
[Koscielny-Bunde et al., 1996; Govindan et al., 2001].
Figure 2(a) shows that the FA (and the similar
Hurst and power spectrum methods) may lead to spurious results because
of the presence of trends, yielding a large overestimation of long
range correlations. Figure 2(d) shows the FA and DFA results for the
artificial data used in Fig. 1(d). Although the profile
function suggested the
presence of a trend, the FA and the DFA show no evidence of any trend
(see references [Hu et al., 2001; Vjushin et al., 2001]). Figures 2(b)
and 2(c) show the FA and DFA results for two
typical sites in the Atlantic and Pacific oceans, respectively. Here,
for long time
scales, FA and DFA curves are straight lines with roughly the same
fluctuation exponent $\alpha \sim 0.8$.
This shows that (a) trends do not
falsify the FA result and therefore may be regarded as much less
important than for Prague temperatures, and (b) long range
correlations also occur in SST's. These correlations are stronger
than the correlations in the atmospheric land temperatures,
since the fluctuation exponent $\alpha \sim 0.8$ corresponds to a
correlation exponent $\gamma \sim 0.4$. As in the case of
atmospheric land temperatures [Koscielny-Bunde et al., 1996],
the range of this persistence law seems to
exceed one decade and is possible even longer than the range of the
SST series considered.

In contrast to Prague, there is a pronounced
short-time regime which ends roughly at 10 months.
This
regime can be better revealed by the analysis of the weekly
SST series. Figure 3 shows the FA and DFA results for $4$ sites in the
Atlantic and Pacific oceans. This figure shows that for short times,
the SST exhibits a persistence
which is considerably stronger than both the SST long term persistence and
the atmospheric land temperature persistence. The typical SST
short-time fluctuation exponent is $\alpha \approx 1.2$.
However, in the
northern Atlantic (latitudes
from 30$^o$ to 50$^o$
north) we have found even higher fluctuation exponents. Figure 3(d)
shows the results for a typical site in
the northern Atlantic, yielding $\alpha \approx 1.4$. The fact that
$\alpha$ is above $1$ means that the variance
of the original temperature fluctuations in a time window $s$
increases as $s^{\alpha - 1}$, i.e. as $s^{0.4}$ in the Northern
Atlantic and $s^{0.2}$ in the rest of the oceans for time scales
below $10$ months. This non-stationary behavior must be
contrasted with the atmospheric land temperature fluctuations where the
variance stays constant and the persistence decays with a nearly
universal exponent $\gamma \sim 0.7$. Non-stationary behavior has also
been found in the analysis of marine stratocumulus cloud base height
records [Kitova et al., 2002].

We like to suggest
the following interpretation for the difference in the short-term
persistence between the Northern Atlantic and the rest of the oceans. In the
northern Atlantic, the dominant mode
of interannual variability in the atmospheric circulation is the
North Atlantic Oscillation (NAO) [Hurrell, 1995; Thompson et al., 1998].
This weather phenomenon highly
influences the climate in the eastern part of North America and northern
Europe and is usually characterized by the NAO index which is based
on the normalized difference in sea level pressure between Ponta
Delgada, Azores (26$^o$ W, 38$^o$ N) and Akureyri, Iceland (18$^o$ W,
66$^o$ N). The
NAO index varies from positive values in winters to negative values in
other seasons. During the last twenty years, the NAO index has displayed a
persistent and exceptionally strong positive phase [Hurrell, 1995].
Since the sea level
pressure and the SST are coupled variables, it is likely that the observed
persistence in the NAO index
is also revealed by the greater fluctuation
exponent found in SST's in the same
period.

In order to find how representative the values of the fluctuation
exponents are,
we have studied the distribution of the short- and long-term exponents
for both the Atlantic and
the Pacific ocean. For the long-term exponents, we exclude those sites in the
tropical Pacific region where the El Ni\~no southern oscillation (ENSO) takes
place [Tziperman et al., 1994; Cane et al., 1986].
The reason for this is that ENSO is a cyclic
phenomenon which warms
the east equatorial Pacific ocean every three to six years. This cycle
cannot be
detrended and strongly affects the DFA results on scales between $2$ and $20$
years. At small scales below $2$y, higher order DFA is able to remove
the trend.
At larger scales, well above $20$y, the oscillations cancel each other and the
fluctuations again become dominant. However, for obtaining reliable results on
the scaling above $20$y, we need data covering far more than
$200$y. Those data are
not available, and therefore we cannot specify the long-term exponents in the
ENSO region. Figure 4 shows the results from our fluctuation analysis for a
typical site in the tropical Pacific region, both for the weekly and
the monthly
data. Below $2$y, the exponent is close to $1.2$, and is therefore
similar to the
short-term exponent for the rest of the sites. Above $2$y, the
influence of the
oscillations shows up. First, the exponent crosses over to a larger
value, and
then, above $3$y for DFA1 and above $8$y for DFA5,  crosses over to a very low
value. This effect of oscillations on the DFA analysis was recently described
in [Kantelhardt et al., 2001; Hu et al., 2001].
We expect that at much larger scales, the exponent
will gradually increase approaching
the value $\alpha \sim 0.8$ as for the sites outside the ENSO
regime. However, the data sets are too short to observe this effect.

Figure 5 summarizes our results for the short- and long-term exponents for both
the Atlantic and the Pacific oceans. As said before, for the short-term
exponents,
sites in the ENSO region are included in the histogram,
while for the
long-term exponents they are not. The histogram shows that the short term
exponents for the
Northern Atlantic ($\alpha = 1.38 \pm 0.04$) where the NAO takes place,
are well distinct from the short term exponents of the remaining sites ($\alpha
= 1.17 \pm 0.08$). For the asymptotic long-term exponents ($\alpha = 0.8 \pm
0.08$) there is not such a
clear distinction between the Northern Atlantic area and the rest.

\section{Conclusions}
In summary, we have studied the persistence of the sea surface temperature in
 the Atlantic and Pacific oceans. We
found that, in
contrast to land stations, there exist two pronounced scaling
regimes. In the short-time regime that roughly ends at 10 months, the
fluctuations of the temperature profile, in a given time window $s$,
scales as $s^\alpha$, with an exponent $\alpha$ in the northern
Atlantic ($\alpha \sim 1.4$) that differs from the rest of the oceans
($\alpha \sim 1.2$). This behavior is well distinct from the
temperature fluctuations on land, where $\alpha$ is close to $0.65$
above typically 10 days. The fact that in the short-time regime
$\alpha$ is well above $1$ points to an intrinsic non-stationary
behavior where the variance of the original temperature fluctuations
in a time window of size $s$ increases with $s$ as $s^{\alpha -
1}$.
This non-stationary behavior crosses over to stationary behavior
at time scales above $10$ months, where now the fluctuation exponent
reaches the value $\alpha \sim 0.8$ for all sites considered in both
oceans. This result reveals that pronounced long term correlations govern the
SST, with an exponent $\gamma \sim 0.4$.
The persistence in the SST is due to the
capacity of the oceans to store heat [Levitus et al., 2001]. The oceans also contribute to
the temperature persistence on land but in a less direct way, i.e. by coupling to the
atmosphere. This may be the reason why the persistence of
atmospheric land temperatures is
less pronounced. In the view of our results, it is interesting that
coastline stations (like Melbourne, Sidney, and New York) show the same
persistence exponent like inland stations (like Prague and Luling).
Finally, we also like to emphasize that the scaling laws we find here may
serve as further non-trivial test-bed for the state-of-the-art global
climate models (see [Govindan R., 2002]).

{\bf Acknowledgments}:
We like to acknowledge financial support from CONICET
(Argentina), the Israel Science Foundation and the Deutsche
Forschungsgemainschaft.

\newpage
\begin{figure}
\noindent\includegraphics[width=39pc]{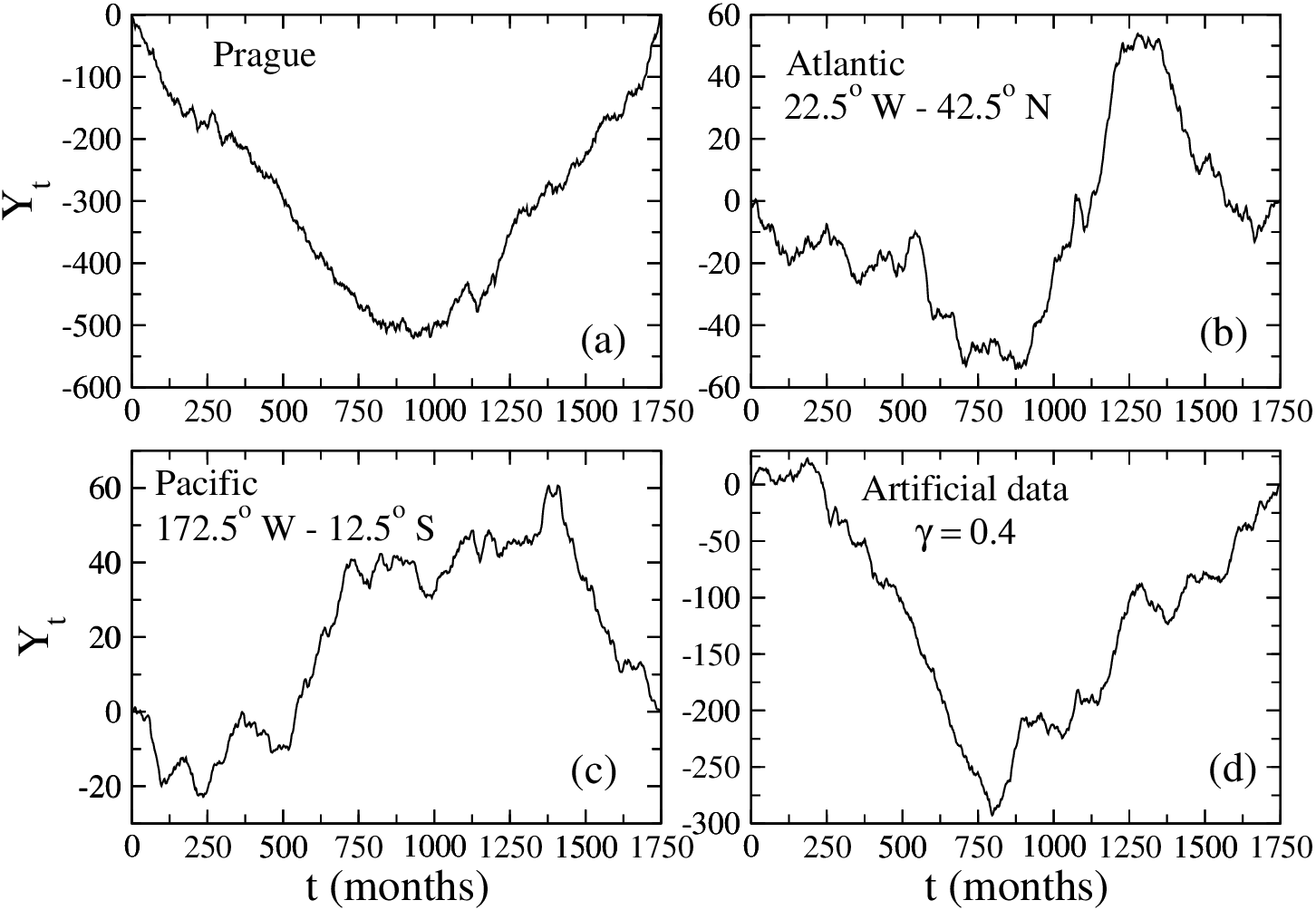}
\caption{Typical temperature profile functions for the last 146 years (monthly
data). (a) Prague, (b) Atlantic ocean (22.5W, 42.5S), (c) Pacific ocean
(172.5W, 12.5S), (d)
Artificial correlated data with $\gamma = 0.4$.}
\label{fig1}
\end{figure}

\begin{figure}
\noindent\includegraphics[width=39pc]{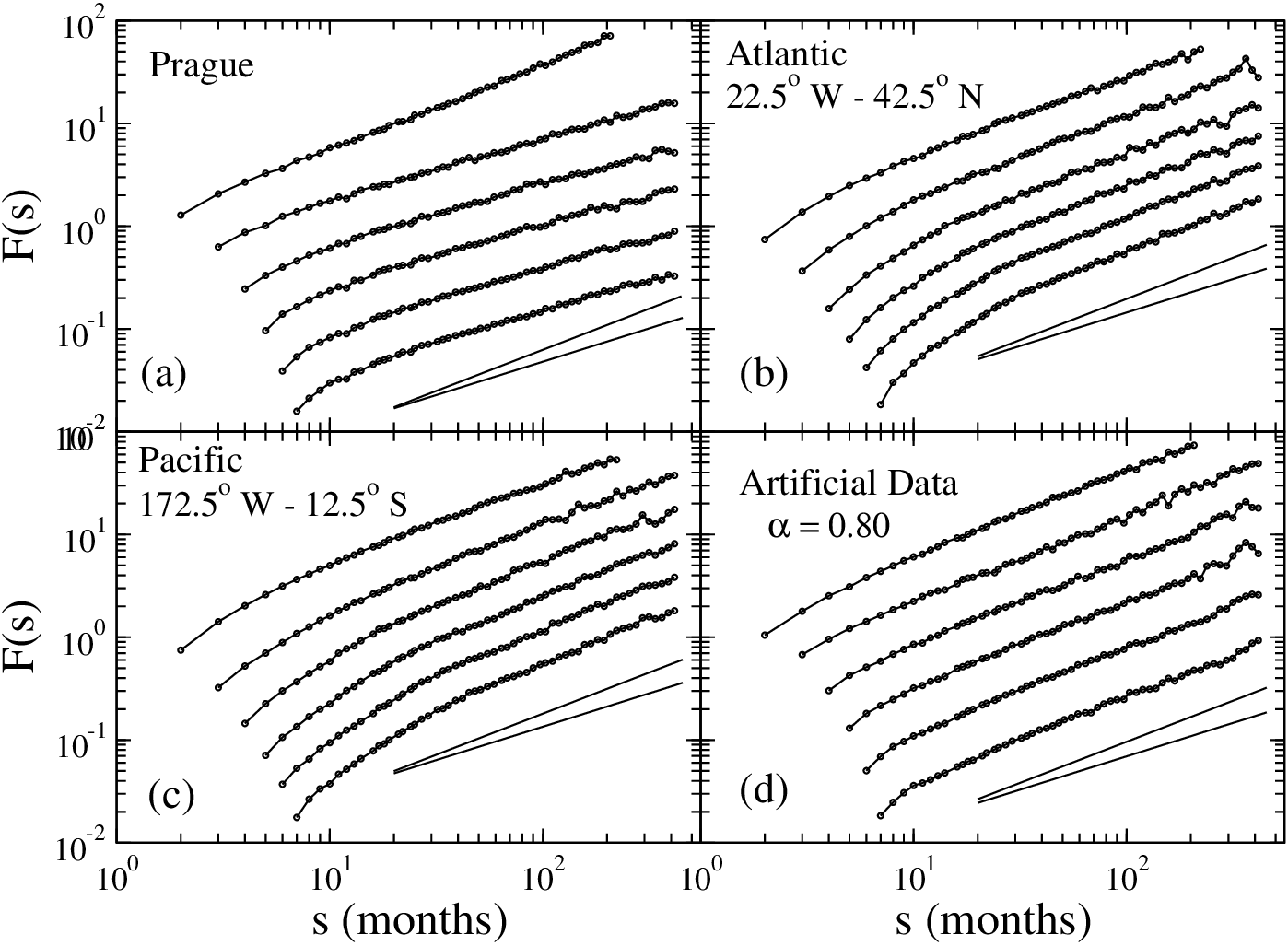}
\caption{Log-log plots of the FA and DFA curves for the data shown in
Fig. 1. From top to bottom curves correspond to FA, DFA1 to
DFA5. Lines of slope $0.8$ and $0.65$ have been drawn to compare the
typical SST long term
fluctuation exponent with the atmospheric land temperature fluctuation
exponent.}
\label{fig2}
\end{figure}

\begin{figure}
\noindent\includegraphics[width=39pc]{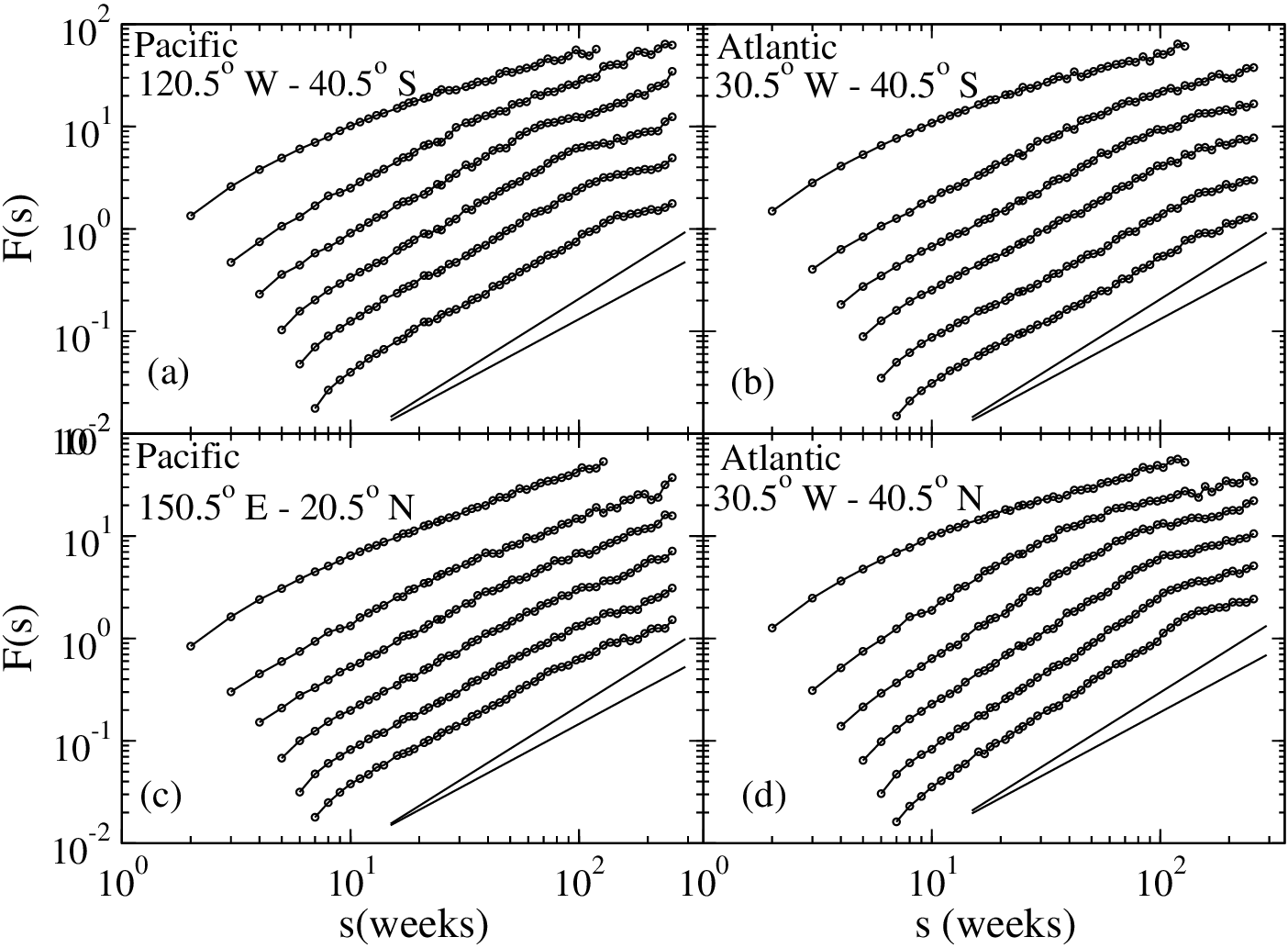}
\caption{Log-log plots of the FA and DFA curves for the
last 20 years (weekly data) for typical sites in the Atlantic and
Pacific oceans. From top
to bottom curves correspond to FA, DFA1 to DFA5. Lines of slopes $1.2$
and $1.4$ have been drawn to compare the short-time SST fluctuation
exponent obtained in the northern Atlantic with the short-time SST
fluctuation exponent for the rest of the oceans.}
\label{fig3}
\end{figure}

\begin{figure}
\noindent\includegraphics[width=39pc]{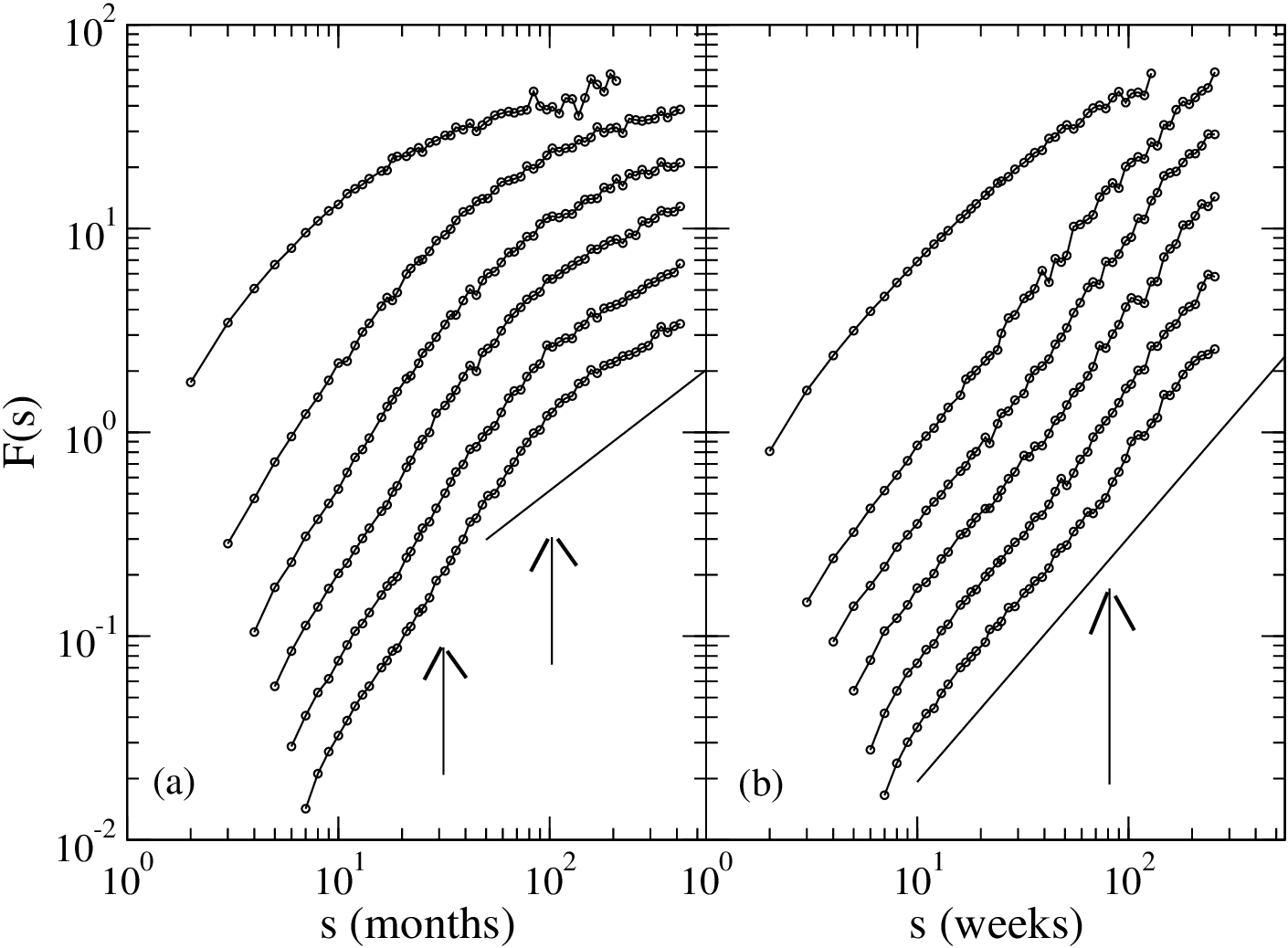}
\caption{Log-log plots of the FA and DFA curves at 92.5$^o$W - 2.5 $^o$S in the
tropical Pacific region. The arrows indicate the position of the
crossovers. (a) Monthly SST's for the last 146 years. A line of slope
$0.8$ has been drawn to note the influence of the oscillation on the
results. (b) Weekly SST's for the last 20 years. A line of slope $1.2$
representative of the short-time regime has been included.}
\label{fig4}
\end{figure}

\begin{figure}
\noindent\includegraphics[width=39pc]{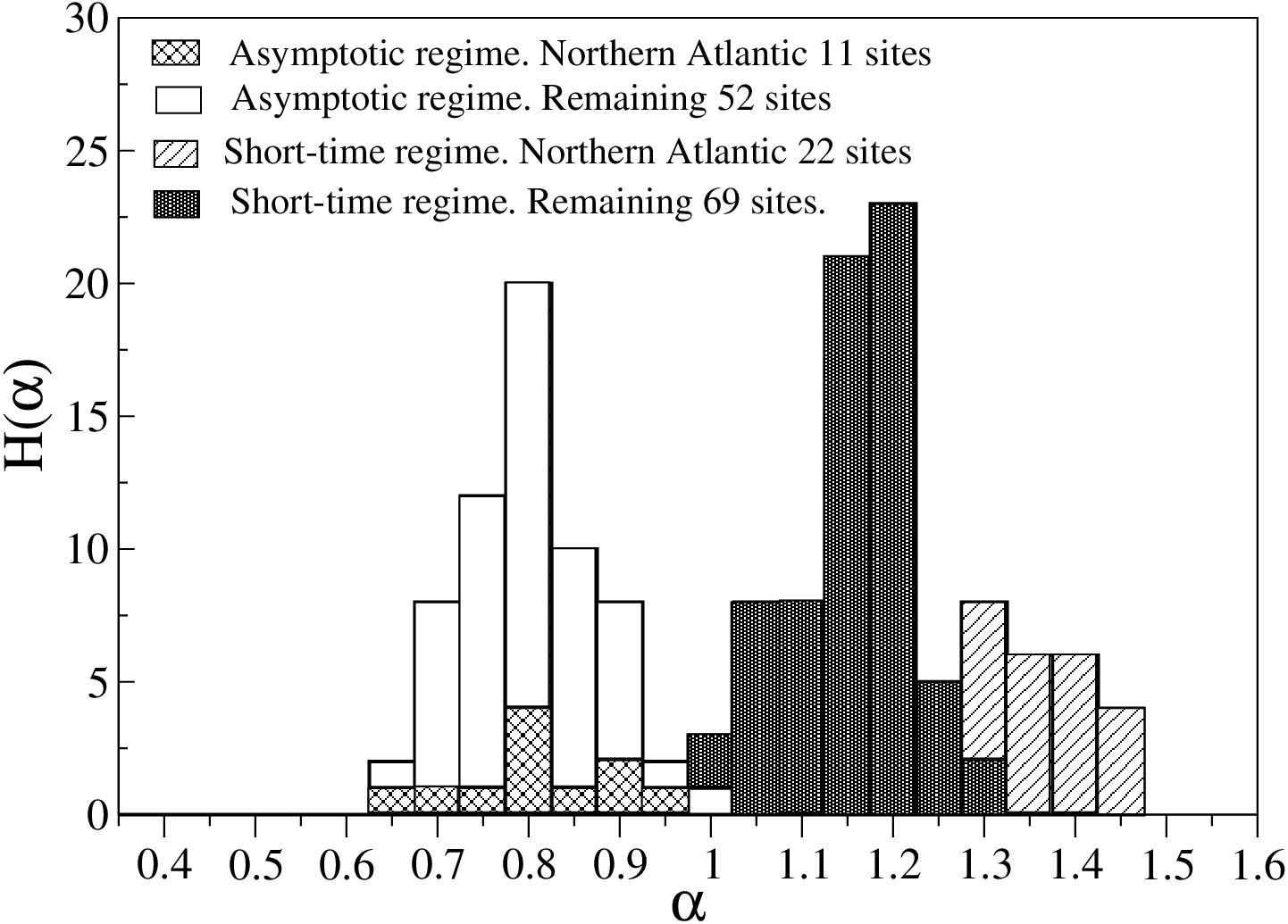}
\caption{Histograms for the short-time and long-time
fluctuation exponents.}
\label{fig5}
\end{figure}


\begin{thebibliography}{}
\bibitem{eva1}
Koscielny-Bunde E., Bunde A., Havlin S., Roman H. E.,
Goldreich Y., Schellnhuber H.-J., Indication of a universal persistence
law governing atmospheric variability, {\it Phys. Rev. Lett., 81,} 729--732,
1998.
\bibitem{} Koscielny-Bunde E., Bunde A., Havlin S., Goldreich Y., Analysis of daily
temperature fluctuations, {\it Physica A, 231,} 393--396, 1996.
\bibitem{pe1} Pelletier J. D., Analysis and modeling of the natural variability
of climate, {\it J. Climate, 10,} 1331--1342, 1997.
\bibitem{pe2} Pelletier J. D. and Turcotte D. L., {\it J. Hydrology,
203,} 198--208, 1997.
\bibitem{talk1} Talkner P. and Webber R.O., Power spectrum and detrended fluctuation
analysis: Application to daily temperatures, {\it Phys. Rev. E, 62,}
150--160, 2000.
\bibitem{Peng1}
Peng C.-K., Buldyrev S.V., Havlin S., Simons M., Stanley H.E.,
Goldberger A.L., Mosaic Organization of DNA Nucleotides, {\it Phys. Rev. E, 49,}
1685--1689, 1994.
\bibitem{Jan}
Kantelhardt J.W., Koscielny-Bunde E., Rego H.H.A., Havlin S.,
Bunde A., Detecting long-range correlations with detrended fluctuation analysis,
{\it Physica A,  295,} 441--454, 2001.
\bibitem{Kap} Kaplan A., Cane M., Kushnir Y., Clement A.,
Blumenthal M., and Rajagopalan B., Analyses of global sea surface temperature 1856-1991,
{\it J. of Geophys. Res-Oceans,  103,}
18567--18589, 1998.
\bibitem{Park} Parker D. E., Jones P. D., Folland C. K., and Bevan A.,
Interdecadal changes of Surface-Temperature since the late-19th-century,
{\it J. of Geophys. Res-Atmos,  99,} 14373--14399, 1994.
\bibitem{Rey1} Reynolds R. and Marsico D., An Improved Real-Time Global Sea-Surface
Temperature Analysis, {\it J. Climate, 6,}
114--119, 1993.
\bibitem{Rey2} Reynolds R. and Smith T., Improved Global Sea-Surface Temperature
Analyses using Optimum Interpolation,
{\it J. Climate,  7,} 929--948, 1994.
\bibitem{rw} Barabasi A.-L and Stanley H. E., {\it Fractal Concepts in
Surface Growth} (Cambridge University Press, 1995).
\bibitem{rw1} Shlesinger M., West B., and Klafter J., Levy Dynamics of Enhanced
Diffusion - Application to Turbulence,
{\it Phys. Rev. Lett., 58,} 1100--1103, 1987.
\bibitem{Govind1}
Govindan R. , Vjushin D., Brenner S., Bunde A. , Havlin S., and
Schellnhuber H.-J., Long-range correlations and trends in global climate models:
Comparison with real data, {\it Physica A,  294,} 239--248, 2001.
\bibitem{Hu} Hu K., Ivanov P.Ch., Chen Z., Carpena P., and
Stanley H. E., Effect of trends on detrended fluctuation analysis,
{\it Phys. Rev. E,  64,} 011114, 2001.
\bibitem{Vjus} Vjushin D., Govindan R., Monetti R., Havlin S.,
and Bunde A., Scaling analysis of trends using DFA, {\it Physica A,
302,} 234--243, 2001.
\bibitem{aus} Kitova N., Ivanova K., Ausloos M., Ackerman T., Mikhalev
M., Time dependent correlations in marine stratocumulus cloud base
height records, {\it Int. Jour. Mod. Phys. C, 13,} 217--227, 2002.
\bibitem{hur} Hurrell J., Decadal Trends in the North-Atlantic Oscillation - Regional
Temperatures and Precipitation, {\it Science,  269,} 676--679, 1995.
\bibitem{thomp} Thompson D. and Wallace J., The Arctic Oscillation signature in the
wintertime geopotential height and temperature fields,
{\it Geophys. Res. Lett.,  25,} 1297--1300, 1998.
\bibitem{tz1} Tziperman E., Stone L., Cane M., and Jarosh H.,
{\it Science, 264,} 72--74, 1994.
\bibitem{} Cane M., Zebiak S., and Dolan S., Experimental Forecast of El-Nino,
{\it Nature, 321,} 827--832, 1986.
\bibitem{} Levitus S, Antonov JI, Wang JL, Delworth TL, Dixon KW, and
Broccoli AJ, Anthropogenic warming of Earth's climate system,  {\it
Science, 292,} 267--270, 2001.
\bibitem{gov2}
Govindan R., Vjushin D., Brenner S., Bunde A., Havlin S.,
Schellnhuber H.-J., Global climate models violate scaling of the observed
atmospheric variability, {\it Phys. Rev. Lett.,  89,} 028501, 2002.
\end{thebibliography}
\end{document}